\documentclass[conference]{IEEEtran}

\usepackage[letterpaper, left=1in, right=1in, bottom=1in, top=0.75in]{geometry}
\usepackage{cite,amssymb}
\usepackage[nolist]{acronym}
\usepackage{graphicx}
\usepackage[cmex10]{amsmath}
\usepackage{subfigure}
\usepackage{color}
\usepackage{multirow}
\usepackage[normalem]{ulem}
\usepackage{soul}
\usepackage{relsize}

\usepackage{amsmath}
\usepackage{algorithm}
\usepackage{algpseudocode}
\makeatletter
\def\BState{\State\hskip-\ALG@thistlm}
\makeatother

\def\Bset[#1]{\mathcal{B}_{#1}}
\def\Vset[#1]{\mathcal{V}_{#1}}
\def\Nset[#1]{\mathcal{N}_{#1}}

\def\Uset[#1]{\mathcal{U}_{#1}}
\def\Kset[#1]{\mathcal{K}_{#1}}

\def\Oset[#1]{\mathcal{S}_{#1}}
\def\Fset[#1]{\mathcal{F}_{#1}}

\usepackage{mathtools}

\newcommand{\density}{\rho}

\newcommand{\Ptx}{P_\text{t}}
\newcommand{\fb}{f_\text{b}}

\newcommand{\Tsense}{\tau_\text{sense}}
\newcommand{\Tbusy}{T_\text{busy}}
\newcommand{\Ttx}{T_\text{tx}}
\newcommand{\ta}{t_\text{a}}
\newcommand{\tb}{t_\text{b}}
\newcommand{\CR}{\omega_\text{CR}}
\newcommand{\CBR}{\delta_\text{CBR}}
\newcommand{\Nsubframes}{S_\text{sense}}
\newcommand{\NoccupSlots}{N_\text{busy}}
\newcommand{\Nused}{S_\text{tx}}
\newcommand{\Nsubchannels}{M_\text{tot}}
\newcommand{\Nsf}{S^*}
\newcommand{\Nneighbors}{V}
\newcommand{\NsubchannelOneTx} {M_\text{p}}

\begin{acronym} 
\acro{3GPP}{Third Generation Partnership Project}
\acro{5GAA}{5G Automotive Association}
\acro{AWGN}{additive white Gaussian noise}
\acro{BEP}{beacon error probability}
\acro{BP}{beacon periodicity}
\acro{BSM}{basic safety message}
\acro{BSS}{basic service set}
\acro{ccdf}{complementary cumulative distribution function}
\acro{cdf}{cumulative distribution function}
\acro{CA}{collision avoidance}
\acro{CAM}{cooperative awareness message}
\acro{CAMs}{cooperative awareness messages}
\acro{CBR}{channel busy ratio}
\acro{C-ITS}{cooperative-intelligent transport systems}
\acro{CR}{channel occupation ratio}
\acro{CSD}{cyclic shift diversity}
\acro{CSI}{channel state information}
\acro{CSMA/CA}{carrier sensing multiple access with collision avoidance}
\acro{C-V2X}{cellular vehicle-to-anything}
\acro{D2D}{device-to-device}
\acro{DCC}{distributed congestion control}
\acro{DCF}{distributed coordination function}
\acro{DCM}{Dual carrier modulation}
\acro{DMRS}{demodulation reference signal}
\acro{DSRC}{dedicated short range communication}
\acro{EDCA}{enhanced distributed coordination access}  
\acro{eNB}{evolved NodeB}
\acro{FCD}{floating car data}
\acro{FD}{full duplex}
\acro{FDD}{frequency division duplex}
\acro{FEC}{forward error correction}
\acro{GNSS}{global navigation satellite system}
\acro{GPS}{global positioning system}
\acro{HD}{half duplex}
\acro{IBE}{in-band emission}
\acro{IPG}{Inter-packet gap}
\acro{ITS}{Intelligent Transportation System}
\acro{i.i.d.}{independent identically distributed}
\acro{KPI}{key performance indicator}  
\acro{LDPC}{low density parity check}
\acro{LOS}{line-of-sight}
\acro{LTE}{long term evolution}  
\acro{LTE-D2D}{\ac{LTE} with \ac{D2D} communications}  
\acro{LTE-V2V}{\ac{LTE}-vehicle-to-vehicle}
\acro{LTE-V2X}{\ac{LTE}-\ac{V2X}}
\acro{LTE-D2D}{long term evolution with device to device communications}  
\acro{MAC}{medium access control}
\acro{MCS}{modulation and coding scheme}
\acro{MIMO}{multiple input multiple output}
\acro{MRD}{maximum reuse distance}
\acro{NGV}{Next Generation V2X}
\acro{NHTSA}{National Highway Traffic Safety Administration}
\acro{NLOS}{non-line-of-sight}
\acro{NR}{new radio}
\acro{NR-V2X}{\ac{NR}-\ac{V2X}}
\acro{OBU}{on board unit}
\acro{OCB}{outside of the context of a \acl{BSS}}
\acro{OFDM}{orthogonal frequency division multiplexing}
\acro{OFDMA}{orthogonal frequency division multiple access}
\acro{pdf}{probability density function}
\acro{PAPR}{peak to average power ratio}
\acro{PD}{packet delay}
\acro{PDR}{Packet delivery ratio}
\acro{PEP}{pairwise error probability} 
\acro{PHY}{physical}
\acro{PL}{path loss}
\acro{ProSe}{proximity-based services}
\acro{QAM}{quadrature amplitude modulation}
\acro{QoS}{quality of service}
\acro{RBP}{resource block pair}
\acro{RR}{radio resource}
\acro{RSU}{road side unit} 
\acro{SC-FDMA}{single carrier frequency division multiple access}
\acro{SCI}{sidelink control information}
\acro{SHINE}{simulation platform for heterogeneous interworking networks}
\acro{SINR}{signal to noise and interference ratio}
\acro{SNR}{signal to noise ratio}
\acro{SPS}{semi-persistent scheduling}
\acro{SRS}{sounding reference signal}
\acro{STBC}{space time block codes}
\acro{SV}{smart vehicle}
\acro{TB}{transport block}
\acro{TDD}{time division duplex}
\acro{TTI}{transmission time interval}
\acro{UE}{user equipment}
\acro{URLLC}{ultra reliable and low latency communications}
\acro{UTDOA}{uplink time difference of arrival}
\acro{V2C}{vehicle-to-cellular} 
\acro{V2N}{vehicle-to-network}
\acro{V2I}{vehicle-to-infrastructure} 
\acro{V2P}{vehicle-to-pedestrian}
\acro{V2R}{vehicle-to-roadside}
\acro{V2V}{vehicle-to-vehicle} 
\acro{V2X}{vehicle-to-everything} 
\acro{WAVE}{wireless access in vehicular environment}
\end{acronym}

\begin{document}

	%
	\title{Congestion Control Mechanisms in\\IEEE 802.11p and Sidelink C-V2X}
	%
	%
	%
	
	
	\author{
		\IEEEauthorblockN{Alessandro Bazzi}
		\IEEEauthorblockA{Universit\`{a} di Bologna-DEI and CNR-IEIIT, Italy \\
			Email: alessandro.bazzi@unibo.it}
		\vspace*{-0.8cm}
	}

	\maketitle
	
	\begin{abstract}
	\footnote{© 2019 IEEE.  Personal use of this material is permitted.  Permission from IEEE must be obtained for all other uses, in any current or future media, including reprinting/republishing this material for advertising or promotional purposes, creating new collective works, for resale or redistribution to servers or lists, or reuse of any copyrighted component of this work in other works.\\Accepted version, presented at Asilomar 2019. }	Connected vehicles are expected to play a major role in the next future to improve safety and traffic efficiency on the road and short-range technologies have been defined to enable the direct exchange of information. To this aim, two solutions are currently the subject of a debate that goes beyond the technician, i.e., IEEE 802.11p and sidelink cellular-vehicle-to-anything (C-V2X). Tested and mature for deployment the first, possibly more efficient the second. In both cases, one of the main aspects is the management of channel congestions, which can cause serious packet losses and have a critical impact on the reliability of applications. Congestions can be managed through different approaches, including the control of transmission power, packet generation frequency, and the adopted modulation and coding scheme. Congestion management has been well studied in IEEE 802.11p, with consolidated algorithms included in the standards, whereas it appears somehow as a new topic looking at C-V2X. In this work, a review of the main congestion control mechanisms and a discussion of their applicability and efficiency in the two technologies is provided. This topic is addressed without focusing on specific algorithms and with the aim to provide general guidelines as a starting point for new proposals.		
	\end{abstract}
	
	
	
	
	\IEEEpeerreviewmaketitle
	
	\acresetall
	
	\section{Introduction}
	
	The automotive sector is expected to be one of the key players of the robot revolution that will radically change our society in the future. Automation and wireless communications will be both crucial in this process and are both subject of huge efforts in industrial and academic research. 
	Regarding the topic of connected vehicles, one of the main aspects is short-range direct communications, enabling improved safety and manoeuvre coordination under the paradigm of \ac{URLLC}. 
	
	In this scenario, two wireless technologies have been developed for \ac{V2X} communications, often referred to as IEEE~802.11p and sidelink \ac{C-V2X} \cite{BazCecMenMasZan:J19}. The former, relatively old and well consolidated, is an adaptation of the classical Wi-Fi, with enhancements currently under definition as IEEE~802.11bd \cite{NaiChoPar:J19}. The latter, at an earlier stage regarding large scale testing, but pushed by a strong ecosystem, was firstly defined in the \ac{LTE} framework as \acused{LTE-V2X}\ac{LTE-V2X} and will be also part of 5G as \ac{NR-V2X} \cite{NaiChoPar:J19}. Given the tremendous market potential and the fact that the same technology seems required on board of all vehicles (at least on a regional basis) to make them inter-operate, a hard discussion is ongoing, which involves all the interested stakeholders, including national governments and international bodies.
	
	The two technologies, briefly discussed in Section~\ref{Sec:technologies}, are viewed by the most as possible alternatives at the lower layers of a general (and common) protocol pillar, defined by SAE and IEEE in the US and by ETSI in Europe \cite{Fes:J2015}. However, they are sensibly different, especially at the \ac{MAC} layer, and thus considerations that were consolidated looking at IEEE~802.11p might not hold moving to sidelink \ac{C-V2X}. One of the aspects that appears not yet fully investigated is the management of \ac{DCC}.
	
	The main traffic carried out by short-range communications, especially in the first phase of V2X deployment, will be the periodic broadcasting by each vehicle of messages detailing its status and movements. We will refer to such messages as beacons in the further to be agnostic to the standards.\footnote{They are part of the \acp{BSM} defined by SAE and correspond to the \acp{CAM} standardized by ETSI.} By its nature, such traffic is proportional to the number of vehicles on the road and tends to saturate the available channel resources when the traffic density increases. A saturation of the channel implies an increase of packet losses and thus a thread for the safety of drivers and passengers.		
			
	Congestion control algorithms have thus been defined to modify some parameters before the conditions become critical \cite{SmeRuhSchKenSjo:BC15}. Several options have been considered for this scope, among which the main ones are acting on the transmission power, beacon frequency, and data-rate. Years of research focusing on IEEE 802.11p have brought to the definition of standard algorithms, based on some continuous measurements performed by the nodes \cite{SAE_DSRC_J2945_1,ETSI_102_687}. Differently, looking at C-V2X, even if a first solution has been defined for example in \cite{ETSI_103_574}, this topic has only gained attention very recently and few early works are available in the literature \cite{ManMarHar:C19,KanJunBah:C18,HaiHwa:J19,SepGozLuc:J19}. 
	
	Rather than looking at specific algorithms, the objective of this work is to overview the main possible measurements and the main adaptable metrics and investigate their role and impact in the two technologies, to provide general indications for the future proposals.	
	
	\section{Main short-range wireless technologies}\label{Sec:technologies}
	
	In this section, the main characteristics of both technologies are shortly recalled. 
	
	\subsection{The Wi-Fi standards}
	
	IEEE 802.11p was firstly defined in 2010, with few modifications compared to the IEEE 802.11a version of Wi-Fi. In particular, it includes: 1)~half bandwidth to reduce noise and improve resilience to channel selectivity; 2)~the so-called \ac{OCB} mode to allow communications without the need of long authentication and association procedures. Later, it was included in IEEE~802.11-2016 \cite{IEEE80211_2016}. In Europe, it goes under the name of ITS-G5.
 	
	At the PHY and MAC layers, it is based on \ac{OFDM} and \ac{CSMA/CA}, respectively. Each message is sent when the channel is sensed idle and occupies the full bandwidth of approximately 10~MHz with a raw data-rate ranging between 3 and 27~Mb/s, depending on the adopted \ac{MCS}. The relatively old PHY layer relies on convolutional coding. The random access scheme at the MAC layer allows easy implementation due to no need of coordination, but is known to have critical risks of congestions and to suffer of strong impact of hidden terminals. Another aspect often criticized to IEEE 802.11p is that the \ac{QoS} cannot be granted.
	
	IEEE 802.11p has been tested in very large implementations and is currently considered a mature technology.	However, in order to update it with the latest solutions, a new working group has been settled in 2018 to define the new IEEE~802.11bd \cite{NaiChoPar:J19}. The main objective is to add advanced coding schemes and other innovations (mainly at PHY layer) to improve the reliability, especially at high speed and with long packets. IEEE~802.11bd is promised to be backward compatible with IEEE~802.11p.
	
	\subsection{The cellular standards}
	
Sidelink LTE-V2X was defined by 3GPP as part of \mbox{C-V2X} in Release 14, with a first publication in 2016 and the frozen version in 2017. It is based on the \ac{D2D} mechanisms, earlier introduced for public safety applications, with enhancement at PHY layer and in the allocation procedures to cope with the specific nature of data traffic and nodes mobility. 
	
	At PHY and MAC layers, it uses the same \ac{SC-FDMA} as LTE uplink. At the PHY layer, a large number of \acp{MCS} based on turbo coding are foreseen, which promise better performance compared to IEEE 802.11p \cite{AnwKulAugFraFet:C18}. At the MAC layer, time and frequency are organized in orthogonal resources that require sophisticated allocation algorithms in order to optimize the spatial reuse. More specifically, the time granularity is given by the sub-frame, or \ac{TTI}, which lasts 1~ms, whereas in the frequency domain the minimum allocation unit is the subchannel, which includes a number of subcarriers and occupies a multiple of 180 kHz. Given the size of the packet and the adopted \ac{MCS}, transmissions are performed using one or more subchannels in one TTI. The allocation procedure can be either controlled by the network (a.k.a. controlled, or Mode~3) or autonomously carried out by each vehicle (a.k.a. distributed, or Mode~4). Since the former relies on an infrastructure, which is not required by IEEE 802.11p, only Mode~4 will be considered in the following. For the details about the Mode~4 algorithm defined by 3GPP, which is based on sensing before select and \ac{SPS}, the reader is referred for example  to \cite{MolGoz:J17,BazCecZanMas:J18}. It is anyway relevant to note that the transmission delay in sidelink \mbox{C-V2X} is constrained by the chosen selection window and that wrong allocations are prone to consequent collisions for several consecutive transmissions.
	
	The cellular ecosystem considers \ac{LTE-V2X} as the baseline for short-range \ac{V2X} communications and the upcoming  \ac{NR-V2X} will be mostly based on similar principles, with the addition of flexible numerology and enhancements for non-periodic traffic. \ac{NR-V2X} is anyway imagined as a separate technology for advanced use cases and retro-compatibility is not addressed \cite{NaiChoPar:J19}.
	
	\section{Distributed congestion control mechanisms}
	
	This section focuses on the main measured metrics and the main parameters that can be modified by \ac{DCC}. 
	
	\subsection{Measured metrics}\label{subsec:measuredmetrics}
	
Given the awareness service and the broadcast nature of the wireless medium, each vehicle has continuous knowledge about how many vehicles are within its transmission range, how much the channel is busy, and how much of this occupation is due to transmissions from itself. These three metrics can all be at the basis of the \ac{DCC} process.
	
	\subsubsection{Channel busy ratio $\CBR$}

	In both IEEE 802.11p and LTE-V2X, each vehicle has the ability to sense the medium in order to estimate when the channel can be used for transmission. Applying this continuously and taking into account the own transmissions, each node can estimate the portion of resources that have been sensed busy in a given time interval, usually known as \ac{CBR}. The CBR is normally at the basis of any DCC algorithm.
	
	\noindent In particular, in IEEE 802.11p the sensing procedure is intrinsically part of the \ac{CSMA/CA} protocol. Given a time interval of length $\Tsense$ and denoting the portion of time between $\ta$ and $\tb$ in which the medium has been sensed busy $\Tbusy(\ta,\tb)$ and the portion of time between $\ta$ and $\tb$ in which the node has been transmitting $\Ttx(\ta,\tb)$, the channel busy ratio $\CBR^\text{11p}(t)$ in an instant $t$ can be calculated as
	\begin{eqnarray}
	\CBR^\text{11p}(t) = \frac{\Tbusy(t-\Tsense,t)+\Ttx(t-\Tsense,t)}{\Tsense}\;.
	\end{eqnarray}
	
	\noindent In LTE-V2X, the power is measured in each subchannel/TTI slot (hereafter \textit{slot}) and compared to a threshold to determine if that slot has been used or not and to estimate what will be the situation in the next subframes. Thus, denoting as $\Nsubframes$ the number of subframes in the given time interval of duration $\Tsense$, $\Nsubchannels$ the number of subchannels, $\NoccupSlots(\ta,\Nsf)$ the number of slots sensed as busy in the $\Nsf$ subframes preceding the instant $\ta$, $\Nused(\ta,\Nsf)$ the number of subframes in which the node was transmitting during the $\Nsf$ subframes preceding the instant $\ta$, the channel busy ratio $\CBR^\text{LTE}(t)$ in an instant $t$ can be calculated as
	\small
	\begin{eqnarray}\label{eq:CBRLTE}
	\CBR^\text{LTE}(t) = \frac{ \NoccupSlots(t,\Nsubframes) + \Nused(t,\Nsubframes)  \cdot \Nsubchannels }{\Nsubframes \cdot \Nsubchannels}\;.
	\end{eqnarray}
	\normalsize
	In \eqref{eq:CBRLTE}, the second term at the numerator takes into account that current devices are half duplex: since a node cannot estimate the use of a slot in a subframe during which its transceiver is in the transmission state, all the slots of such subframes are assumed busy.	
	
	\subsubsection{Number of neighbors $\Nneighbors$}

	Each vehicle continuously receives the periodic beacons from the neighbors and updates a list of neighboring nodes with all information related to their position and movements. The number of neighbors that can be inferred from the list is a metric used by some algorithms in addition to the \ac{CBR}.
	
	\subsubsection{Channel occupation ratio $\CR$}

	Each vehicle can calculate how much of the resources is itself occupying. Such metric, normally called \ac{CR}, is calculated as the portion of time/bandwidth used in a given time window. If the periodic transmission of beacons is the only traffic, then it can be calculated in IEEE 802.11p as 
	\begin{eqnarray}
	\CR^\text{11p}(t) = \tau_\text{p}^\text{11p}(t)\cdot \fb(t)
	\end{eqnarray}
	where $\tau_\text{p}^{11p}(t)$ is the duration of a transmission and $\fb(t)$ is the beacon frequency at the time of calculation.

	Similarly, in LTE-V2X it can be calculated as 
	\begin{eqnarray}
	\CR^\text{LTE}(t) = \left(\frac{\NsubchannelOneTx (t)}{\Nsubchannels}\tau_\text{TTI}\right)\fb(t)
	\end{eqnarray}
	where $\NsubchannelOneTx(t)$ is the number of subchannels used to transmit a packet at the time of calculation and $T_\text{TTI}$ is the duration of a TTI (1~ms in LTE, variable in NR).

	\subsection{Modified parameters}
	
	A number of parameters can be modified in order to impact on the channel occupation. 
	Among them, those that are mostly considered are power, packet rate generation, and data-rate \cite{SmeRuhSchKenSjo:BC15,PatShe:C18}. 
	
	\subsubsection{Transmission power $\Ptx$} Intuitively, if the transmission power is reduced, the interference and the message losses due to collisions can be mitigated. The expected cost is a reduction of the transmission range. 
	
	\subsubsection{Beacon frequency $\fb$} A lower beacon frequency reduces the channel occupation. The probability of collisions thus reduces without altering the transmission range. The drawback in this case is a lower update frequency of the information regarding the position and movements of the neighboring vehicles.
	
	\subsubsection{MCS} Changing the MCS impacts on the trade-off between data-rate and reliability. Reducing the data-rate, the reliability increases if the interference is not considered. However, transmissions occupy more resources and might increase the probability of concurrent transmissions and consequent collisions. 

	\section{Performance}
	
	
	In this section, results are shown from simulations performed using the LTEV2Vsim simulator~\cite{CecBazMasZan:C17}. 
	
	\subsection{Simulation configuration} 

\begin{table}
\caption{Main simulation parameters and settings.
\vspace{-2mm}
\label{Tab:Settings}}
\scriptsize
\centering
\begin{tabular}{p{4.5cm}p{2.6cm}}
\hline
\textbf{\textit{Common settings}} & \\
Beacon frequency ($\fb$)  & 10 Hz (*) \\
Beacon size & 300~B \\
Time interval for CBR assessment ($\Tsense$) & 1 s\\
MCS & B of Table \ref{Table:LTEMCS} (*) \\
Channel bandwidth & 10~MHz \\
Transmission power ($\Ptx$) & 23~dBm (*) \\
Antenna gain (both tx and rx)  & 3 dB \\
Noise figure & 9~dB \\
Propagation model & WINNER+, Scenario B1 \\
Shadowing & Variance 3 dB, \\
& Decorrel. distance 25~m \\
\textbf{\textit{Related to IEEE 802.11p}} & \\
Duration of the initial interframe space & 110\;$\mu$s \\
Random backoff & $[0\div 15] \cdot 13$\;$\mu$s\\ 
Carrier sensing threshold & -65 dBm \\
\textbf{\textit{Related to LTE-V2X (Mode 4)}} & \\
Probability to maintain the allocation ($p_\text{k}$) & 0 \\
Sensing threshold to assume the channel busy & -94~dBm \\
\hline
(*) used if not differently specified.&\\
\end{tabular}
\end{table}

\begin{table}[t!]
\caption{Considered MCSs. Minimum SINR with both technologies; duration of a packet of 300 bytes with IEEE~802.11p*; packets of 300 bytes per TTI (1~ms) with LTE-V2X. \vspace{-2mm}
\label{Table:LTEMCS}}
\centering
\begin{tabular}{p{0.5cm}p{1.6cm}p{1.1cm}p{0.9cm}p{1.4cm}}
\hline
\textbf{MCS} & \textbf{Mod./Coding}  &
\textbf{Min SINR} & \textbf{Duration 802.11p*} & \textbf{Packets/TTI \mbox{LTE-V2X}}\\  
A & QPSK, 0.27 & 1.49~dB & 560~$\mu$s & 1\\ 
B & QPSK, 0.48 & 5.79~dB & 304~$\mu$s & 2 \\ 
C & 16QAM, 0.46 & 12.83~dB & 192~$\mu$s & 3  \\ 
D & 16QAM, 0.59 & 16.39~dB & 160~$\mu$s & 4  \\  
\hline
\end{tabular}
\end{table}


\begin{table}[t!]
\caption{Channel occupation ratio (CR) for various settings \vspace{-2mm}
\label{Tab:CR}}
\centering
\begin{tabular}{p{0.5cm}p{1.2cm}p{1.5cm}p{1.5cm}}
\hline
\textbf{MCS} & \textbf{Beacon frequency}  &
\textbf{CR in IEEE 802.11p*} & \textbf{\mbox{CR in} \mbox{LTE-V2X}}\\  
A & 10 Hz & 0.0056 & 0.01 \\ 
B & 10 Hz & 0.003 & 0.005 \\ 
C & 10 Hz & 0.0019 & 0.0033 \\ 
D & 10 Hz & 0.0016 & 0.0025 \\  
B & 5 Hz & 0.0015 & 0.0025 \\ 
B & 1 Hz & 0.0003& 0.0005 \\ 
\hline
\end{tabular}
\vskip -0.3cm
\end{table}


\begin{figure*} [t]
	\centering
    \subfigure[IEEE 802.11p*.]{
	\includegraphics[width=0.4\linewidth,draft=false]{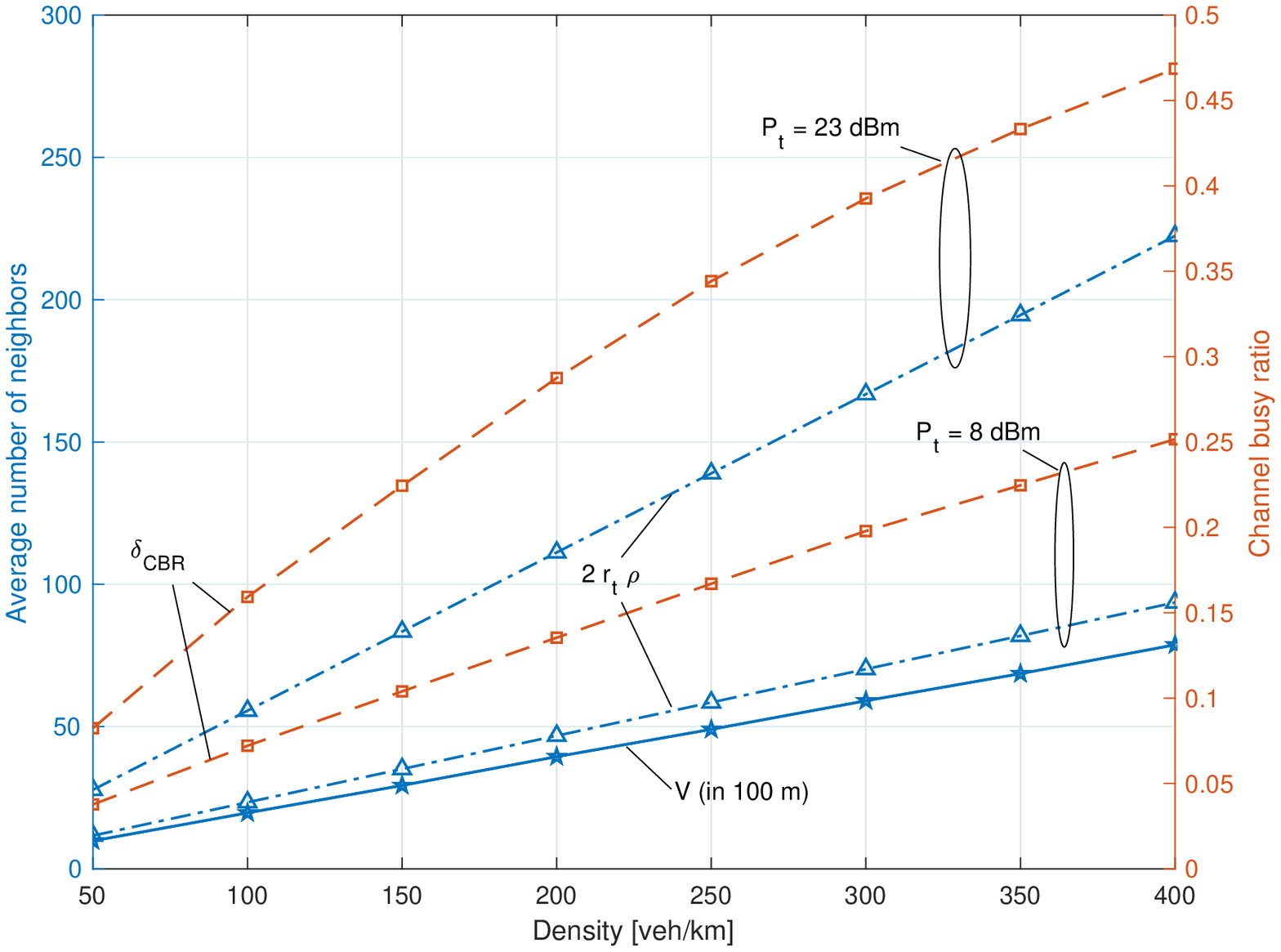}\label{fig:neighb11p}}~~~~
	\subfigure[LTE-V2X.]{
	\includegraphics[width=0.4\linewidth,draft=false]{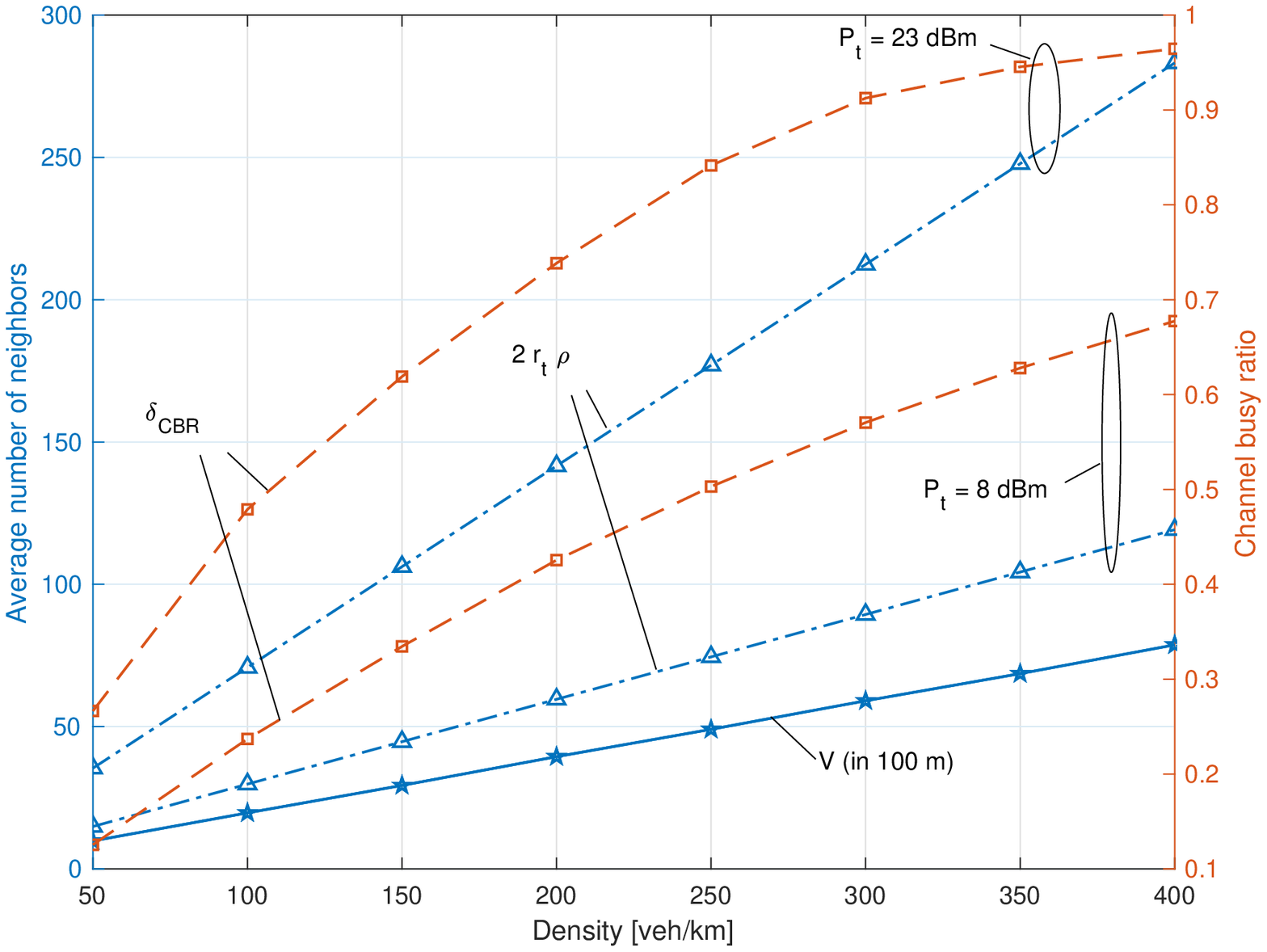}\label{fig:neighbLTE}}		\caption{Average number of neighbors in the median transmission range, average number of neighbors $\Nneighbors$ within 100~m,  and channel busy ratio $\CBR$ vs. density $\density$. Minimum (8~dBm) and maximum (23~dBm) power~$\Ptx$.}
	\label{fig:neighb}\vskip -0.3cm
\end{figure*}	

\subsubsection{Scenario and main settings} In order to reproduce a highway scenario with a variable number of nodes, a 1-D Poisson point process with variable density $\density$ is adopted as for example in \cite{ZhaCheYanEtAl:J12,ParKimHon:J18}. Both technologies are assumed to occupy one 10 MHz channel in the 5.9~GHz frequencies, which have been reserved to this scope in most countries worldwide. The path loss is modeled using the WINNER+ Model B1, as suggested by 3GPP, with correlated shadowing (variance 3~dB, decorrelation distance 25~m). Beacons of 300~bytes are transmitted with a periodicity between 1 and 10~Hz. The transmission power is set between 8 and 23~dBm, plus antenna gains of 3~dB. The noise figure is set to 9~dB. In the case of LTE-V2X, the keep probability adopted by Mode~4 is set to 0 in order to minimize the probability of consecutive collisions 	\cite{MolGozSep:C18,BazCecZanMas:J18}. A summary is provided in Table~\ref{Tab:Settings}.

\subsubsection{IEEE~802.11p*} In order to avoid IEEE 802.11p vs. LTE-V2X performance comparisons and focus on the resource allocation and congestion control, here we assume a modified PHY layer for IEEE 802.11p, which performs similarly to LTE-V2X (same approach as in \cite{BazZanMas:J19}). More specifically, the \ac{CSMA/CA} mechanism, the inter-frame spaces, and the duration of the packet preamble are the same as in IEEE 802.11p; differently, the minimum \ac{SINR} and the raw data rate during transmission are those of LTE-V2X (refer to Table~\ref{Table:LTEMCS}). 

\subsubsection{MCS} The selected MCSs are listed in Table~\ref{Table:LTEMCS}; the values are taken from \cite{BazCecMenMasZan:J19} and correspond to the minimum MCS in LTE-V2X to have one, two, three, or four packets of 300~bytes per TTI. The minimum SINR level is obtained as detailed in \cite{BazZanMas:J19} and the same value is set both in LTE-V2X and IEEE 802.11p*. In IEEE 802.11p*, the duration of a transmission is calculated using the raw data-rate of LTE-V2X obtained as in \cite{BazZanMas:J19}.

\subsubsection{Outputs} The following metrics are considered:
\begin{itemize}
	\item \textit{\ac{PDR}:} it is the ratio between the messages correctly received by the nodes at a given distance and all the nodes at that distance;
	\item \textit{\ac{IPG}:} it is the time between two consecutive correct receptions at a given receiver from the same transmitter, within a given distance; IPG implicitly gives information about the correlation among errors.
\end{itemize}


	\subsection{Results}
\begin{figure*} [t]
	\centering
	\subfigure[Varying $\Ptx$.]{
		\includegraphics[width=0.31\linewidth,draft=false]{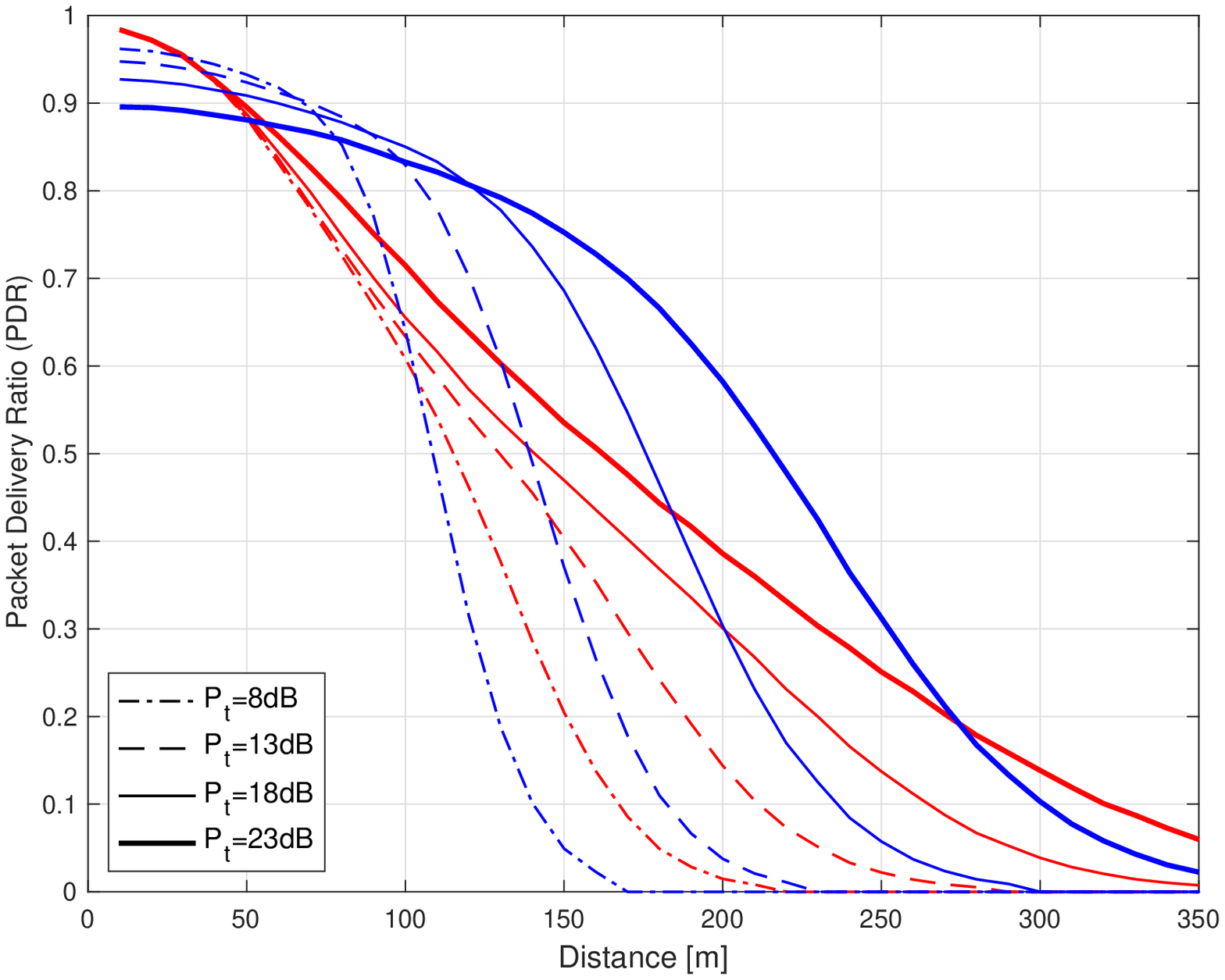}\label{fig:PRRdistanceP}}~~
	\subfigure[Varying $\fb$.]{
		\includegraphics[width=0.31\linewidth,draft=false]{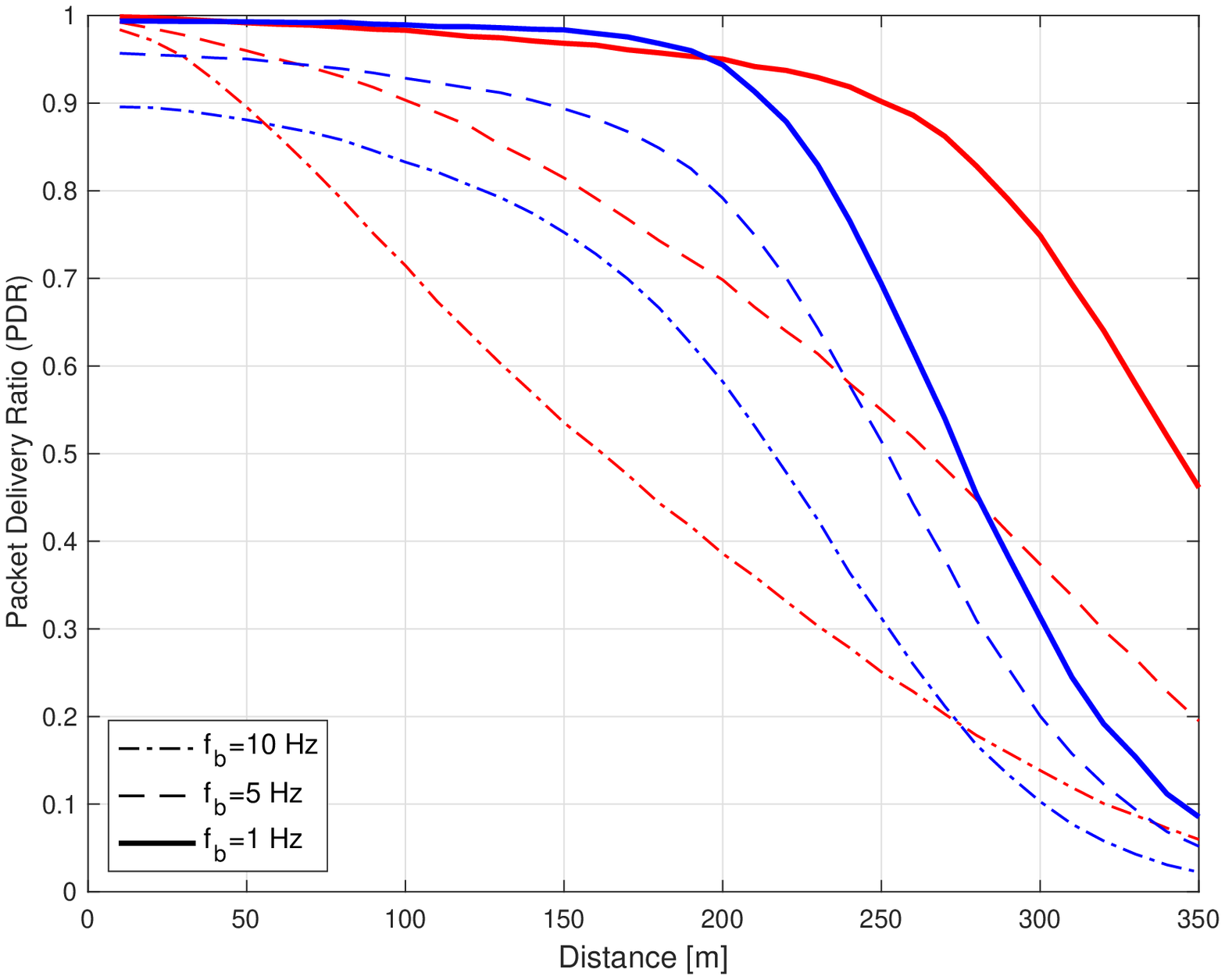}\label{fig:PRRdistanceTb}}
	\subfigure[Varying the MCS.]{
		\includegraphics[width=0.31\linewidth,draft=false]{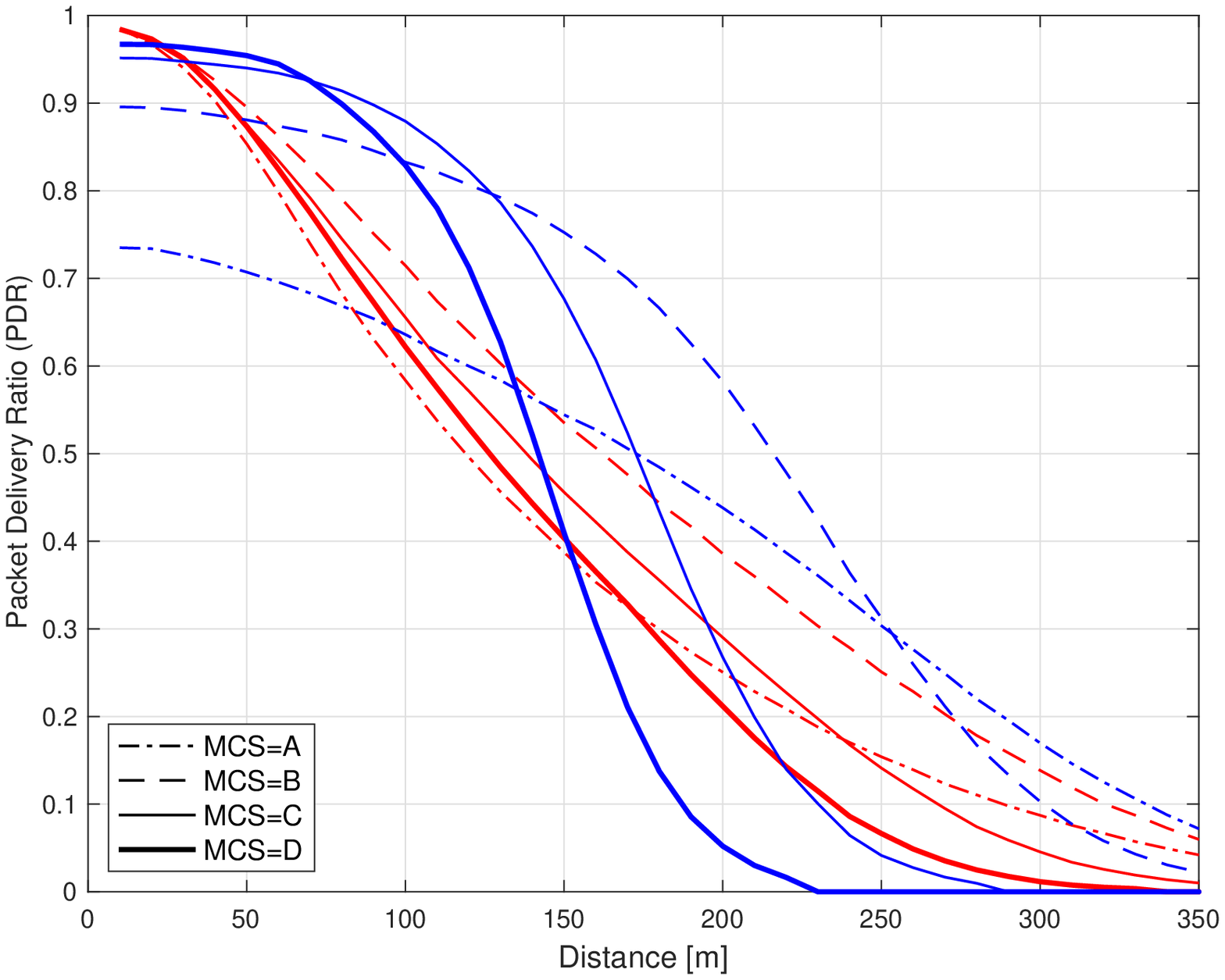}\label{fig:PRRdistanceMCS}}~~
	\caption{Packet reception ratio vs. distance. $\density=300$~vehicles/km. Blue=IEEE 802.11p*, Red=LTE-V2X.}
	\label{fig:PRRdistance}\vskip -0.3cm
\end{figure*}

\begin{figure} [t]
	\centering
	\includegraphics[width=0.62\linewidth,draft=false]{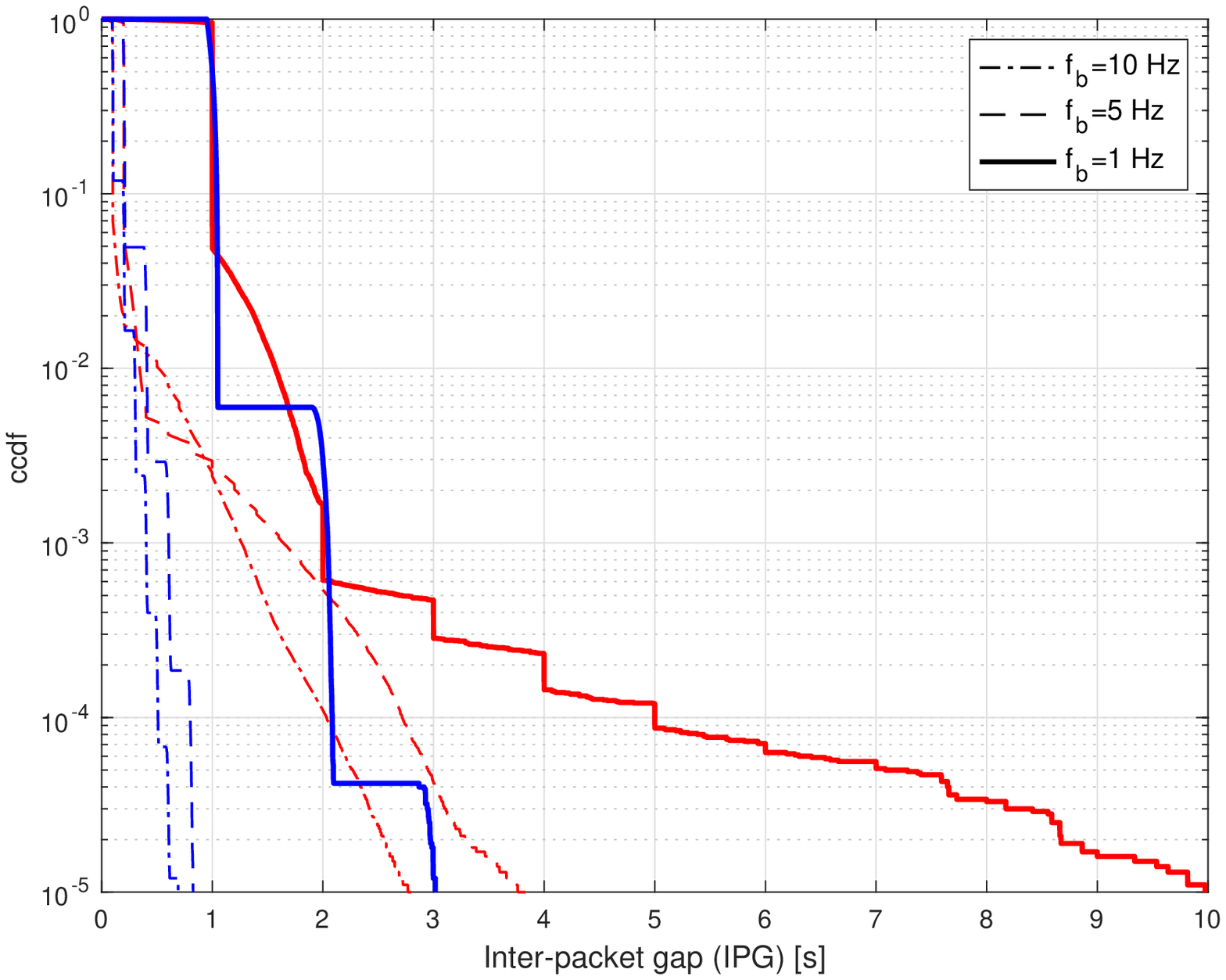}
	\caption{ccdf of the update delay for neighbors within 100~m, varying $\fb$. $\density=300$~vehicles/km. Blue=IEEE 802.11p*, Red=LTE-V2X.}
	\label{fig:UDdistance} \vskip -0.3cm
\end{figure}

	Results are shown through Figs.~\ref{fig:neighb}-\ref{fig:UDdensity}, firstly focusing on the measured metrics and then investigating the performance that derive from parameter variations. The \ac{CR} calculated with the adopted combinations of the parameters is also given in Table~\ref{Tab:CR} for the sake of completeness.
	

	\subsubsection{Observing the metrics measured for DCC} Fig.~\ref{fig:neighb} focuses on the metrics measured by the vehicles and possibly used to determine the level of congestion of the channel. 
	The two subfigures correspond to IEEE 802.11p* (Fig.~\ref{fig:neighb11p}) and LTE-V2X (Fig.~\ref{fig:neighbLTE}), respectively. More specifically, the following three metrics are shown varying the vehicle density $\density$, for both the lowest and highest transmission power: 1) the average number of vehicles that are within the median transmission range, obtained as the multiplication of the  density by the range with path loss only; 2) the average number of neighbors that the vehicles perceive within a range of 100~m; 3) the \ac{CBR}, measured as detailed in Section~\ref{subsec:measuredmetrics}. 

	The average number of vehicles within the median range gives an idea of the scenario and the density perceived by the vehicles. A higher value is observed with LTE-V2X, since half bandwidth is used (thus half noise at the receiver). The other two metrics are those normally used for congestion estimation. Whereas the measured average neighbors within 100~m is linear with the density and do not change varying the transmission power, the measured $\CBR$ depends on the power level and tends to be concave, especially in LTE-V2X with $\Ptx=23$~dBm. With the considered values (see Table~\ref{Tab:Settings}), which are those normally adopted by the standards, the $\CBR$ of LTE-V2X appears significantly higher than that of IEEE 802.11p*. To be noted that, whereas in IEEE~802.11p* the channel sensed busy differs the transmission and a high $\CBR$ might cause large delays and even starvation, in LTE-V2X the measurements are only used to identify the preferable TTI/subchannel combinations and do not have any impact on the average delay.
	
	\subsubsection{Impact of power variations} Figs.~\ref{fig:PRRdistanceP} and~\ref{fig:PRRdensityP} show the effect of power variations in both technologies. As observable, especially through Fig.~\ref{fig:PRRdistanceP}, its variation is quite effective in IEEE 802.11p* to trade-off channel congestion and range. Reducing the power, the coverage reduces, but the probability that the channel is sensed as idle by a node increases and the spatial reuse of resources also increases. 
	Indeed, power variations are for example at the core of  SAE J2945/1 \cite{TogSaiFalMug:C19}.
	
	Differently, power adaptation is ineffective in LTE-V2X and using the maximum value appears almost always preferable. This is a consequence of the fact that all signals are synchronized and that sensing is used only to select the 20\% least interfered slots: if the performance is limited by noise, a reduction of the transmission power reduces the signal to noise ratio, while  if the performance is limited by interference, a reduction of the transmission power from all nodes does not alter the signal to interference ratio. These results find confirmation in related work. Ineffectiveness of power variations is in fact shown in \cite{TogSaiFalMug:C19}, where the same SAE J2945/1 algorithm is applied to the cellular technology. Even in \cite{KanJunBah:C18,HaiHwa:J19}, where improvements varying the power level in LTE-V2X are shown, they appear very limited.
	
	\subsubsection{Impact of packet rate variations} 
\begin{figure} [t]
	\centering
	\includegraphics[width=0.62\linewidth,draft=false]{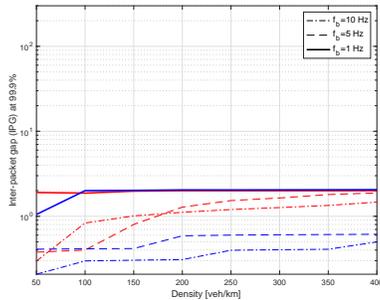}
	\caption{99.9\% of the update delay vs. density for neighbors within 100~m, varying $\fb$. $\density=300$~vehicles/km. Blue=IEEE 802.11p*, Red=LTE-V2X.}
	\label{fig:UDdensity} \vskip -0.3cm
\end{figure}

The performance varying the frequency of beacon generation is shown through Figs.~\ref{fig:PRRdistanceTb}, \ref{fig:UDdistance}, \ref{fig:PRRdensityTb}, and~\ref{fig:UDdensity}. Firstly focusing on the \ac{PDR} (Figs.~\ref{fig:PRRdistanceTb} and \ref{fig:PRRdensityTb}), it can be noted that this approach is effective in both technologies. Actually, its effectiveness is even higher in LTE-V2X, where less packets generated on average corresponds to a higher average reuse distance of the same resources. If we focus as an example on Fig.~\ref{fig:PRRdistanceTb} and LTE-V2X, the maximum distance to have a PDR above 0.9 increases from 25~m to 250~m if $\fb$ is reduced from 10 to 1~Hz.
	
	The beacon periodicity variation is indeed one of the main parameters used in IEEE~802.11p (especially in Europe) and currently the main one considered for LTE-V2X \cite{ETSI_103_574}.
	
	The main drawback of this approach is visible  in Figs.~\ref{fig:UDdistance} and~\ref{fig:UDdensity} observing the \ac{IPG}, which increases sensibly for lower $\fb$. It can also be noted that this issue is more severe in LTE-V2X. This effect is a consequence of the specific resource structure and allocation algorithm of this technology.

	\subsubsection{Impact of data-rate variations} 
\begin{figure*} [t]
	\centering
	\subfigure[Varying $\Ptx$.]{
		\includegraphics[width=0.31\linewidth,draft=false]{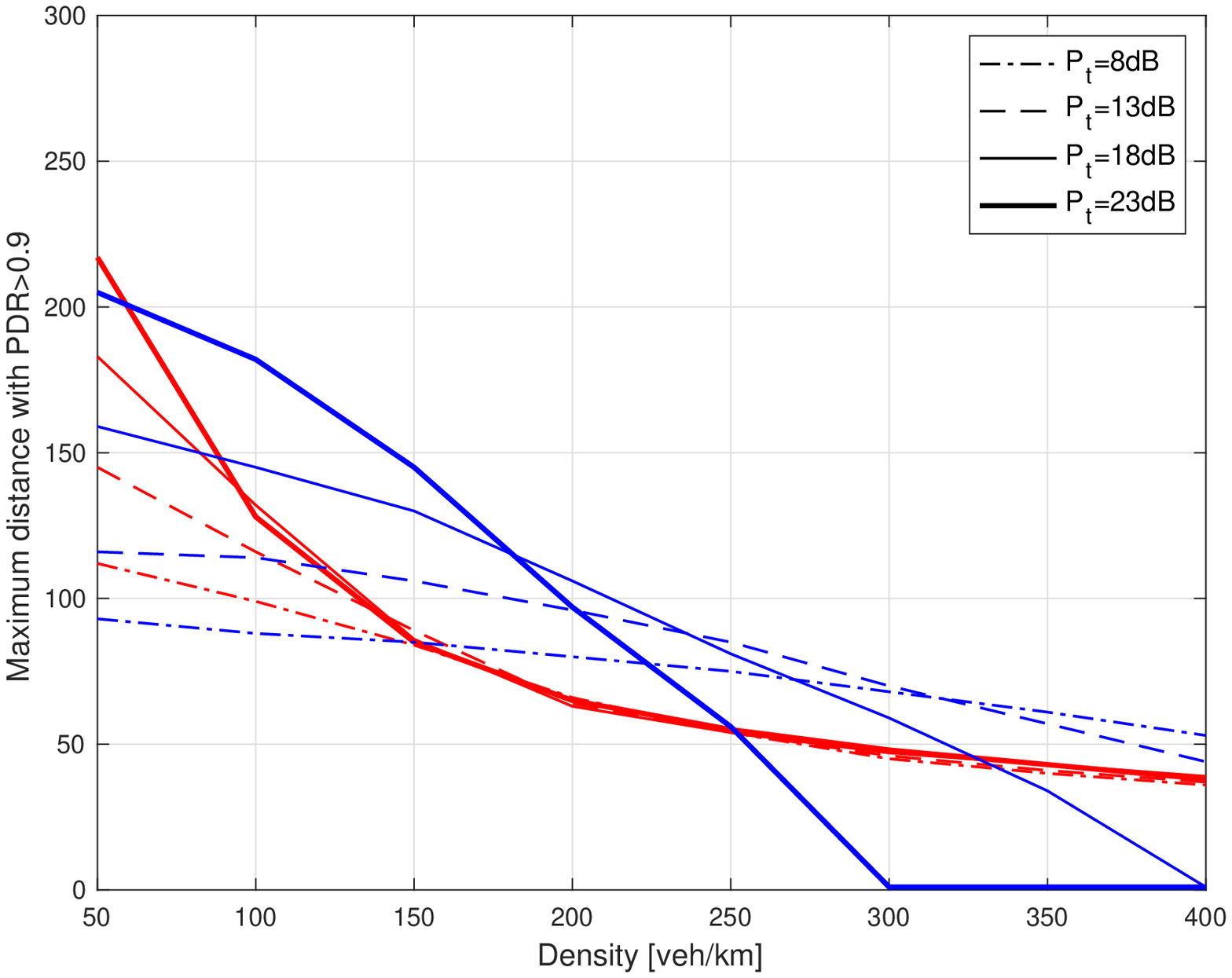}\label{fig:PRRdensityP}}~~
	\subfigure[Varying $\fb$.]{
		\includegraphics[width=0.31\linewidth,draft=false]{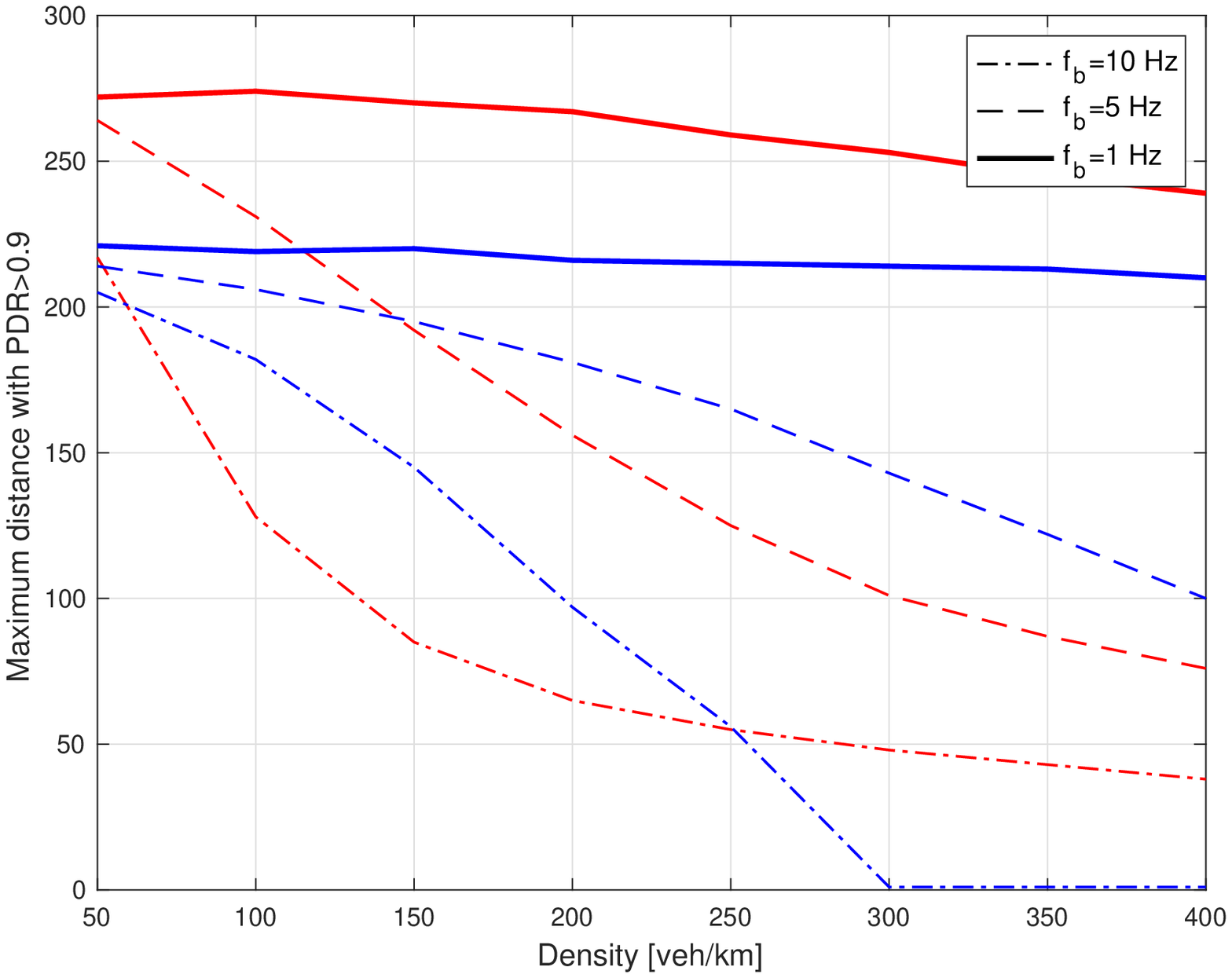}\label{fig:PRRdensityTb}}
	\subfigure[Varying the MCS.]{
		\includegraphics[width=0.31\linewidth,draft=false]{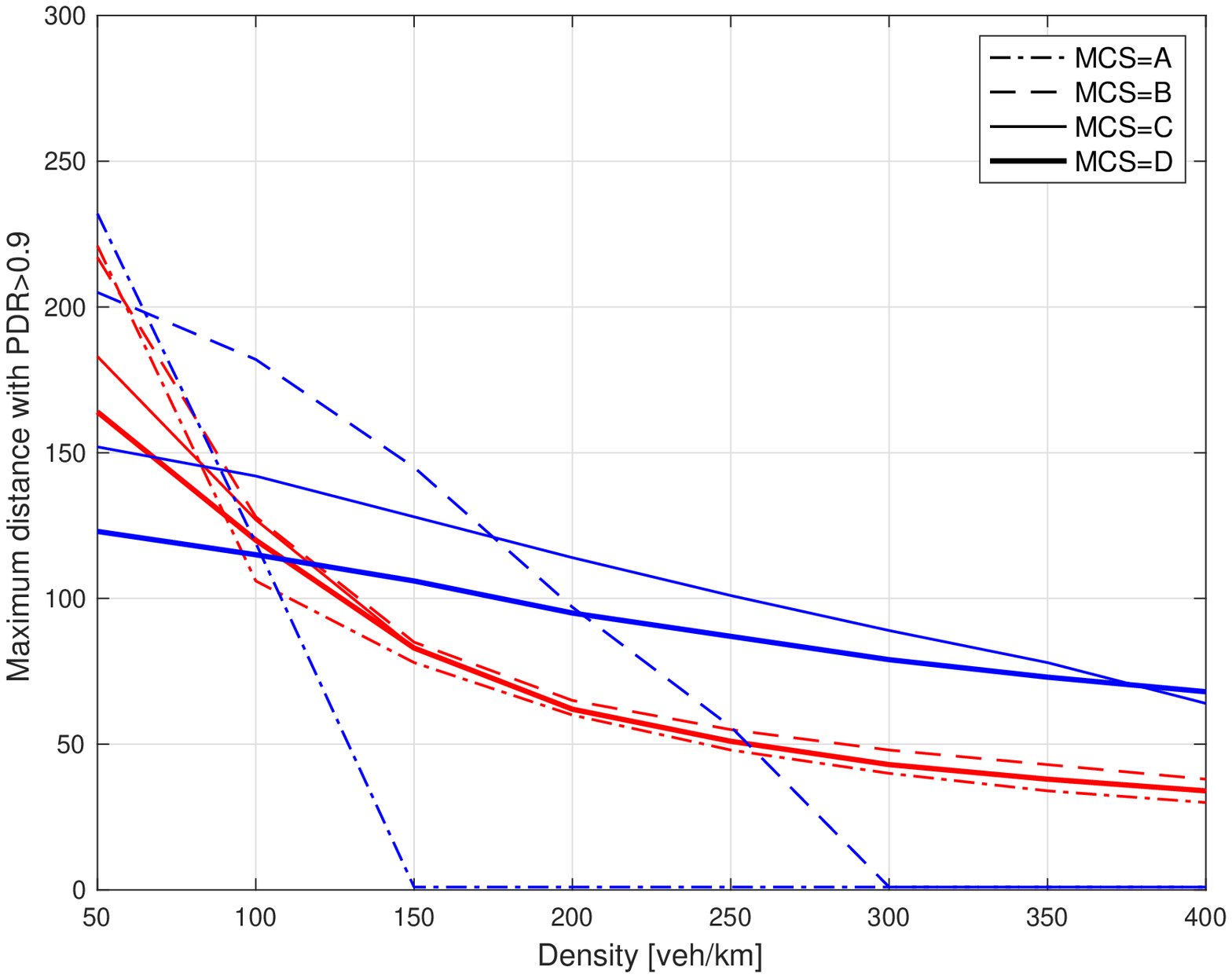}\label{fig:PRRdensityMCS}}~~
	\caption{Packet reception ratio vs. density for neighbors within 100~m. Blue=IEEE 802.11p*, Red=LTE-V2X.}
	\label{fig:PRRdensity}\vskip -0.3cm
\end{figure*}

Finally, the impact of a different selection of the MCS is shown in Figs.~\ref{fig:PRRdistanceMCS} and \ref{fig:PRRdensityMCS}. Similarly to the power variations, also in this case the trade-off between range and congestions is clear in IEEE 802.11p*, whereas it is very limited in LTE-V2X. Moreover, while the optimal MCS depends on both the density and the targeted distance in IEEE 802.11p, it is only related to the density in LTE-V2X.

\section{Conclusion}

In this work, channel congestion control is addressed focusing on both IEEE 802.11p and sidelink LTE-V2X. In particular, after giving an overlook on the measurements performed by nodes to estimate the level of resources occupation, the impact of a variation of three parameters (transmission power, packet generation rate, and \ac{MCS}) is evaluated in a highway scenario. Results show that in IEEE 802.11p either of the three parameters can be efficiently used to trade-off congestions with range (power and MCS) or delay (packet generation rate). Differently, in LTE-V2X: i) power variations appear basically ineffective; ii) acting on the packet generation rate is effective, but may lead to high delays in awareness update; and iii) the optimal choice of the MCS has small impact and is only a function of the density of vehicles.


\bibliographystyle{IEEEtran}
\bibliography{biblioSelf,biblioOthers,biblioStandards}

\end{document}